\documentclass[preprint,12pt]{elsarticle} 
\usepackage{graphicx} 
\usepackage{amssymb} 
\usepackage{amsmath} 
\usepackage{amsfonts} 
\usepackage{slashed} 
\usepackage[dvipsnames]{xcolor} 
\newcommand{\nn}{\nonumber\\&&}
\newcommand{\ben}{\begin{displaymath}}
\newcommand{\een}{\end{displaymath}}
\newcommand{\be}{\begin{equation}}
\newcommand{\ee}{\end{equation}}
\newcommand{\bea}{\begin{eqnarray}}
\newcommand{\eea}{\end{eqnarray}}

\newcommand{\eq}[1]{Eq.~(\ref{#1})}

\newcommand{\bfk}{{\bf k}}                                                           

    \newcommand {\boldsigma}{\mbox{\boldmath$\sigma$}}

\def\G{\Gamma}
\newcommand{\la}{\langle}
\newcommand{\ra}{\rangle}

\journal{Nuclear Physics A} 

\begin{document}

\begin{frontmatter}

\title{ Luneburg-lens-like structural Pauli attractive core of the nuclear force at short distances
 }  

\author{Gerald~A.~Miller} 

\cortext[cor1]{corresponding author: Gerald A. Miller, miller@uw.edu} 
\address{Department of Physics,
University of Washington, Seattle, WA 98195-1560} 

\begin{abstract} 
A recent paper [S. Ohkubo, Phys. Rev. C 95, 044002 (2017)] found that the  measured $^1S_0$  phase shifts can be reproduced using  a deeply attractive nucleon-nucleon   potential.  We find  that the deuteron would decay strongly  via pion emission  to the  deeply bound state arising in this  potential. Therefore the   success of a  deeply attractive potential in describing phase shifts must be regarded only as  an interesting  curiosity.
 \end{abstract}


\end{frontmatter}

%\newpage 

\section{Introduction} 
\label{sec:intro}

  A recent paper \cite{Ohkubo:2017pmh} finds a  nuclear force with an attractive potential at short distances that reproduces the experimental $^1S_0$ phase shifts well. Such a potential can be motivated by early quark-model ideas~\cite{Neudatchin1977}, but later work ~\cite{Liberman1977} showed that quark model ideas  lead to short distance repulsion between nucleons.  Here we show that the  deep attraction causes  a deeply bound state to exist, with  the drastic consequence that the deuteron would not be  stable. 

The $^1S_0$ potential  $V(r)$ of \cite{Ohkubo:2017pmh}  is given by 
\bea&&
 V(r)  =  -5\, e^{-(r/2.5)^2} -270\, e^{-(r/0.942)^2}  -1850\, e^{-(r/0.447)^2}\nn
 \equiv \sum_{n=1}^3 V_n e^{-r^2/r_n^2}.
\label{V}
\eea
The strength  parameters of $V$ are given in units of MeV, and range  parameters are in units of  fm.
This purely attractive potential has a depth of 2125 MeV at $r=0$ and a half-width $r_0$ of about 0.4 fm. The corresponding uncertainty principle estimate of the kinetic energy, $\hbar^2/(M  r_0^2)$, with $M$ as the nucleon mass, is 259 MeV,  so that  the quickest look at this potential leads to the conclusion that  the existence of a deeply bound state  is an immediate  consequence of using \eq{V}. 

\begin{figure}[h]
\centering
\includegraphics[scale=0.3]{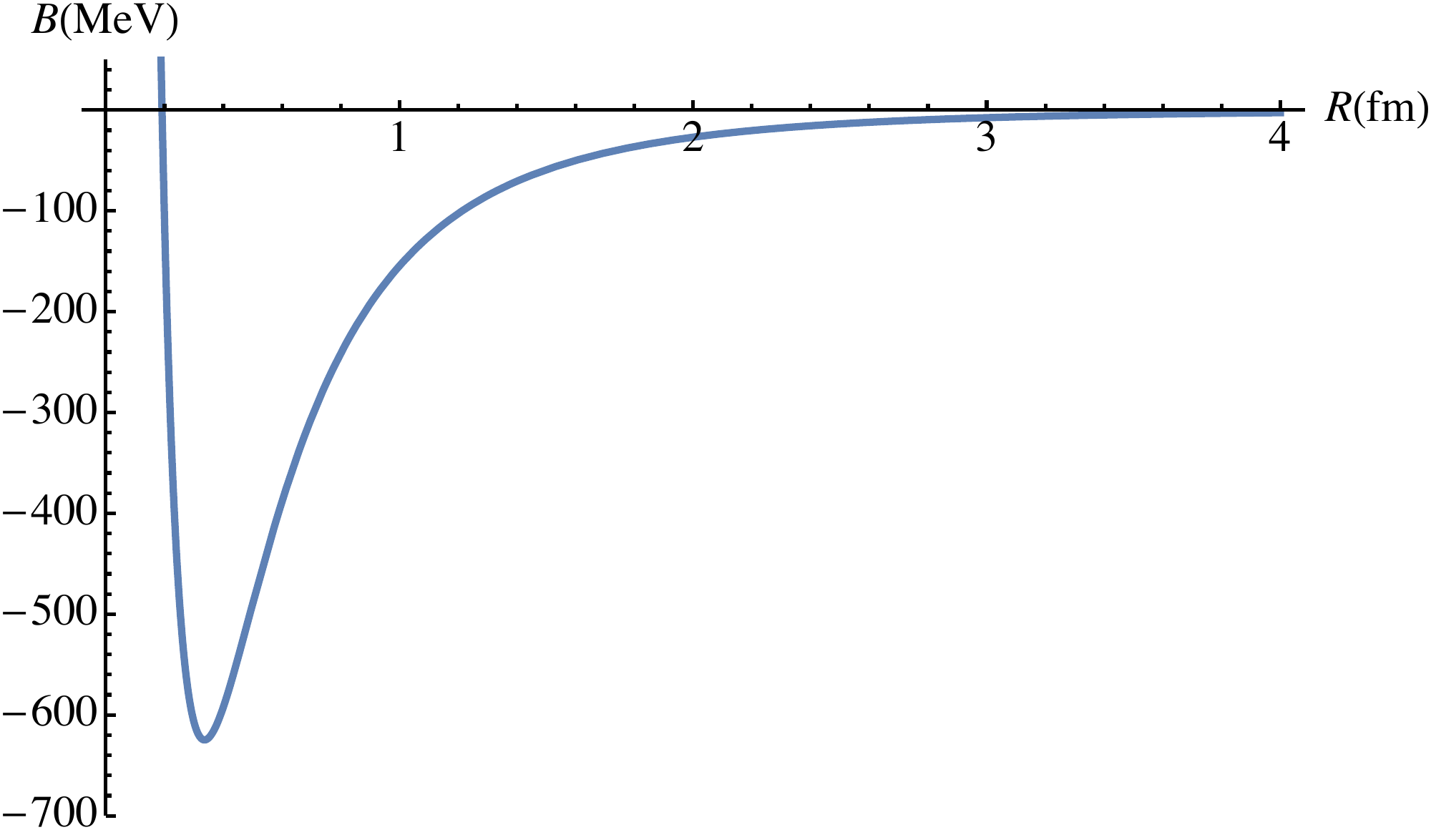}
\caption{\label{fig} (Color online)  The expectation value of the Hamiltonian, $B(R)$.   }
\end{figure}

 The easiest analytic way to show that a bound state must exist is to use the variational principle. The single-parameter trial wave function $u(r)$ used here takes the form:
\bea
u(r)=\frac{2 r e^{-\frac{r^2}{2 R^2}}}{\sqrt[4]{\pi } \sqrt{R^3}},
\eea
  with the normalization $\int_0^\infty \,dr\, u^2(r)=1$.  If the expectation value of the Hamiltonian, $H$,  within this (or any) wave function is less than zero, the potential must yield a bound state. The expectation value of the $H$, defined as $B(R)$ 
  is given by
 % \end{document}
  \bea B(R) =\frac{3 \hbar}{2 M R^2} \ 
   +\sum_{n=1}^3 V_n { 1\over (1+{R^2\over r_n^2})^{3/2}},\label{bnd}
  \eea
 with the  first, positive term arising from the  kinetic energy much smaller than the negative potential energy terms. This can be seen immediately using only  the $V_3=-1850 $ MeV term of \eq{bnd}. With $R=r_3$ the $V_3$ term is 
 $V_3/(2\sqrt{2})=-650
 $~MeV,  while the kinetic energy term is about 390 MeV.
   Fig.~\ref{fig}  shows that  $\langle H\rangle\equiv B(R)$ bottoms out at about -620  MeV. Thus there must be a bound state, and its binding energy must be greater than or equal to 620 MeV. Numerical solution of the Schroedinger equation yields a binding energy of  about 640 MeV~\cite{Ohkubo:2017pmh}.
  % \cite{Ohkubo:2017pmh}.

In Ref.~\cite{Ohkubo:2017pmh}.
this state is  denoted  as ``unphysical" and ``Pauli forbidden".  However, the Pauli principle does not forbid a $^1S_0$ bound state. For example 6 quarks each in the lowest orbital of the MIT bag model form a bound state in that channel if gluon exchange effects are neglected~\cite{Jaffe:1976yi}
and such states could play an important role in nucleon-nucleon scattering \cite{Liberman1977,Harvey:1988nk}.

 A deeply bound $^1S_0$  state has never been found and our very existence shows that this bound state cannot be real.  This is because the deuteron 
 would decay  strongly to this bound state by the emission of a pion.   
 
 \section{Deuteron instability}
 
 We  proceed to  compute the width  $\G$  and lifetime using first order perturbation theory. 
 The pion ($\pi^+$) emission interaction Hamiltonian, $H_I$ is given in first-quantized notation by,
 \bea 
 H_I={g\over 2M}\sum_{i=1,2}\boldsigma_i\cdot \bfk\tau^-_i,
 \eea
 where the pion nucleon coupling constant, $g$ is taken to be, $g^2/(4\pi)=13.5, \,\bfk$ is the pion momentum in the center-of-mass frame, and the operators $\boldsigma_i,\tau^-_i$ are usual Pauli spin and isospin operators that act on nucleon $i$. The operator $H_I$ connects the initial deuteron state (of spin $m$) with the final two-body  state, with a matrix element
 \bea
 {\cal M}_m\equiv \la B|H_I|D,m\ra,
 \eea
 and the total decay width $\G$ is given (after evaluating the phase space integral)  by
 \bea
 \G={1\over 2\pi} {1\over 3}\sum_m |{\cal M}_m|^2k.
 \eea
Evaluation yields the expression
 \bea \G={2\over 3}{g^2\over 4\pi}{k^3\over M^2}(I_0+{\sqrt{2}\over2}I_2)^2\eea
 where \bea I_l\equiv\int dr u_B(r) u_l(r)j_l(kr/2),\eea
with $u_B(r)  $ is    the radial  $^1S_0$  bound state wave function produced by the potential of \eq{V},  $u_{0,2}$ are  the $S$ and $D$ state deuteron radial functions, and $j_{0,2}$ are spherical Bessel functions.   With  a bound state of 640 MeV, $k=728 $ MeV/c.  Numerical evaluation using the deuteron wave function  of the Argonne V18 potential~\cite{Wiringa:1994wb} gives $I_0=0.0925,\,I_2=0.0161$, so that the net 
 result is $\G= 42  $ MeV, which corresponds to a lifetime $T={\hbar \over \G} =1.6 \times 10^{-23}$ s, so that deuterons could not exist. 
 
 \section{Discussion}
 There are many other possible reactions for which the use of this potential would have drastic erroneous consequences.  Immediate examples are the  transition amplitudes for $np\to d\gamma$ and more importantly the  $pp\to D e^-\bar{\nu}_e$ reaction that is essential for understanding the energy radiated by our sun. 
The low-energy nucleon-nucleon wave functions  of the potential of \eq{V} have a node~\cite{Ohkubo:2017pmh}, which arises from the necessary orthogonality of the bound state with all  scattering states. The nodes in the wave function would vastly reduce the  mentioned  transition amplitudes.

The possibility of a purely attractive nucleon-nucleon potential, with its connection to a Luneburg lens~\cite{lb}  is very interesting. But such an interaction can only be regarded as an oddity  unrelated to the real  Universe.\\

    This work   was supported by the U.S. Department of Energy Office of Science, Office of Nuclear Physics under Award No. DE-FG02-97ER-41014. I thank the Physics Department of the Hebrew University  and the Physics Division of Argonne National Laboratory for their  hospitality during  visits.

\end{document}